\documentclass{article}

\usepackage{amsmath}
\usepackage{amscd}
\usepackage{amsthm}
\usepackage{amssymb} \usepackage{latexsym}
\usepackage{euscript}
\usepackage{epsfig}
\usepackage{tikz}
\usepackage{graphics}
\usepackage{array}
\usepackage{enumerate}
\usepackage{authblk}
\usepackage[colorlinks=true]{hyperref}
\usepackage{authblk}
\usepackage{pdfpages}

\newcommand{\be}{\begin{equation}}
\newcommand{\bel}[1]{\begin{equation}\label{#1}}
\newcommand{\qe}{\end{equation}}

\begin{document}
\title{Phylogeny of Metabolic Networks: A Spectral Graph Theoretical Approach}

\author[1]{\rm Krishanu Deyasi}
\author[1,2]{\rm Anirban Banerjee}
\author[3]{\rm Bony Deb}
\affil[1]{Department of Mathematics and Statistics}
\affil[2]{Department of Biological Sciences}
\affil[ ]{Indian Institute of Science Education and Research Kolkata}
\affil[ ]{Mohanpur-741246, India}
\affil[3]{Department of Biological Science}
\affil[ ]{Presidency University}
\affil[ ]{Kolkata-700073, India}
\affil[ ]{\textit {anirban.banerjee@iiserkol.ac.in, krishanu1102@iiserkol.ac.in, bony.d92@gmail.com}}

\maketitle
\bigskip

\begin{abstract}
Many methods have been developed for finding the commonalities between different organisms to study their phylogeny. The structure of metabolic networks also reveal valuable insights into metabolic capacity of species as well as into the habitats where they have evolved. 
We constructed metabolic networks of 79 fully sequenced organisms and compared their architectures. We used spectral density of normalized Laplacian matrix for comparing the structure of networks. The eigenvalues of this matrix reflect not only  the global architecture of a network but also the local topologies that are produced by different graph evolutionary processes like motif duplication or joining. A divergence measure on spectral densities  is used to quantify the distances between various metabolic networks, and a split network is constructed  to  analyze the phylogeny from these distances. 
In our analysis, we focus on the species, which belong to different classes, but  appear more related to each other in the phylogeny. We tried to  explore whether they have evolved under similar environmental conditions or have similar life histories.
 With this focus, we have obtained interesting insights into the phylogenetic commonality between different organisms.
\end{abstract}

\section{Introduction}
Analyzing the commonalities between biological organisms and classifying them is a fundamental approach in evolutionary studies.
Organisms are primarily divided into three domains: Archaea, Bacteria, and Eukaryote. A traditional way to find phylogenetic
relationships between different species is based on similarity in the sequence of orthologous genes, e.g., SSU rRNA (16S rDNA) 
gene sequence \cite{Olsen94}. It is not easy to identify orthologs and paralogs in the genomes due to gene duplication, gene loss and horizontal gene transfer. 
Moreover in ``molecular approach",
protein sequences are also exploited to study the evolutionary relationships \cite{Doolittle99, Woese98}. However, the effect of
horizontal gene transfer  \cite{Doolittle99, Wolf 02}, which has been observed  in many organisms, on the phylogenetic relationship
among species, considering the system level of different cellular functions, is not very clear. Some studies have shown that more than 10\% 
genes in organisms including bacteria, archaea, and eukaryotes are laterally acquired \cite{Garcia-Vallv00, Hedges02, Koonin01, Ochman00}. 
This supports a strong influence of horizontal gene transfer in cellular evolution \cite{Dutta02, Jain02, Martin99}. 
However, since an organism may have genes from different sources, i.e., may have more than one ancestor,  the evolutionary history of organisms can be better represented by a net rather than a tree \cite{Doolittle99} and \cite{Martin99}. With ever-growing molecular 
data, we have many fully sequenced genomes which are very useful for comparative analysis  of organisms at a system level
and which may give us new insights into the process of evolution of organisms that could not be explored by traditional phylogenetic analyses 
based on a limited number of  genes or proteins \cite{Wolf 02}.

The annotations of the metabolic reactions contain information regarding cellular activities and  their presence in various species \cite{Kanehisa06}.
Distance between species can be estimated based on the content of genes encoding enzymes in their genome, or on the metabolic reactions network,
or both (links between these two aspects have been explained in \cite{Liu07}). A method to construct a phylogeny, based on enzyme, reaction, and gene content comparison can be found in \cite{ma04}. 
A phylogeny could be reconstructed  by using graph kernel for comparing metabolic network structures \cite{Oh06}.
The set-algebraic operations have been used \cite{Forst06} and  the seed compound content has been compared  \cite{Borenstein08} to trace the phylogeny. 
By computing the distances between the vectors of several network-descriptors  estimated on the network of interacting pathways, the cross-species phylogenetic distance could be predicted \cite{Mazurie08}. 
The study of the co-evolutionary relationships, by comparing the metabolic pathways to the evolutionary relationship between different species based on the combined similarities of all of their metabolic pathways, can be found in \cite{Mano10}. 

In this work, we perform extensive phylogenetic comparison of 
79 completely sequenced organisms (7 eukaryotes, 13 archaea, 59 bacteria)
by comparing the structure of their metabolic networks at system level. It is expected that the three domains: Archaea, Bacteria and Eukaryote would be
clearly visible. But we focus on the species which come close to each other in our study in spite of belonging to different classes and attempt to find similarity in their life histories.
However, our method is not able to strongly  demonstrate ``horizontal gene transfer'' as a reason for this closeness.
To investigate the structural similarities  between various metabolic networks, we compare spectral density of the normalized graph 
Laplacian \cite{BanerjeeJost2007} and analyze the biological significance for the species whose spectral density is
similar to each other. The spectrum of the normalized graph Laplacian not only reflects global structure of a network but also the local
topologies that emerge from different graph operations such as, duplication of a motif (induced subgraph), attachment of a small structure 
into the existing network \cite{BanerjeeJost2009, AbRm, Vukadinovic2002} etc. The distribution of the spectrum is considered
as a signature of a network structure, and the spectral plot can distinguish network structure from different sources \cite{BanerjeeJost2008b}.
The networks constructed from the identical evolutionary processes have similarities in their 
spectral plots and thus the spectral distance between two networks can be used to devise a similarity measure of their structures \cite{banerjee12}.
\section{Methods}
\subsection{Normalized Graph Laplacian Spectrum}
Let $G$ be an undirected and unweighted network with the vertex set $V=\{i:i=1,\dots ,N \}$. 
If the vertices $i$ and $j$ are connected by an edge, we call them neighbors and it is denoted by $i\sim j$. The number of neighbours of $i$ is 
called the degree of $i$ and is denoted by $d_i$.

The normalized graph Laplacian ($N\times N$) matrix \cite{BanerjeeJost2008a} is defined as:\\
 $\Delta =(\Delta)_{ij}$, $i,j=1,\ldots,N$ where
\begin{eqnarray}\label{1}
\left(\Delta\right)_{ij} &:=& \left\{
\begin{array}{r cl} 1 & if & i = j \mbox{ }
\\ -\frac{1}{d_i} &if& \mbox{$i\sim j$}
\\0&&  \mbox{otherwise.}
\end{array} \right.
\end{eqnarray}
Note that, this operator is similar with the operator studied in \cite{Chung} but is different than the operator widely studied as
(algebraic) graph Laplacian (see \cite{Mohar1991}). Now we can write the eigenvalue equation as $\Delta u-\lambda u=0$ where the
nonzero  solution $u$ is called an eigenfunction for the eigenvalue $\lambda$. Some of the eigenvalues among $N$ eigenvalues of 
$\Delta$ may occur with higher (algebraic) multiplicity.

The eigenvalues of $\Delta$ are real and non-negative and the smallest eigenvalue is always $\lambda_1=0$ for any constant 
eigenfunction $u$. The multiplicity of eigenvalue zero reflects the number of connected components in the network. The lowest
non-zero eigenvalue informs us how easily one graph can be cut into two different disjoint components. A graph is bipartite if
and only if the highest eigenvalue of $\Delta$, $\lambda_{N} = 2$. Moreover, for a bipartite graph the spectral plot is symmetric about 1.  
For a complete connected graph, all non-zero eigenvalues are equal to $\frac{N}{N-1}$ (see \cite{Chung} for the details).

The eigenvalues also reflect the local structures produced by certain graph operations like 
doubling of an induced subgraph (motif) or joining of another graph
\cite{BanerjeeJost2009}. For example, duplication of a single vertex (the simplest motif)
produces an eigenvalue $1$, which is widely observed with higher multiplicity in many biological networks. If we duplicate 
an edge $(i,j)$ (motif of size two), it generates the eigenvalues $\lambda_\pm=1\pm\frac{1}{\sqrt{d_{i}d_{j}}}$ and if 
 a chain $(i-j-k)$ of length $3$ is duplicated, it produces the eigenvalues $\lambda=1,1\pm\sqrt{\frac{1}{d_{j}}(\frac{1}{d_{i}}+\frac{1}{d_{k}})}$.
 With the specific value of the degrees of these vertices,  the above two motif duplications produce the eigenvalues 
 $1\pm0.5$ and $1\pm\sqrt{0.5}$ which are also frequently observed in the spectrum of biological and other networks. 
Joining a small graph $\Gamma$ (with an eigenvalue $\lambda$ and corresponding eigenfunction that vanishes at a vertex 
$i\in\Gamma$) by identifying the vertex $i$ with any vertex of a graph $G$ produces a new graph with the same eigenvalue 
$\lambda$. E.g., joining a triangle (which itself has an eigenvalue $1.5$) to a graph contributes the same eigenvalue $1.5$  
to the new graph produced by the joining process (for more details see \cite{BanerjeeJost2009}).
\subsection{Spectral Density, Network Distance and Clustering of Species}
Now we convolve the spectrum of a network with a kernel $g(x,\lambda)$ and get the function
 \be
f(x)=\int g(x,\lambda)\sum_k \delta (\lambda, \lambda_k)d\lambda= \sum_k g(x,\lambda_k).
\qe
Clearly, $0<\int f(x)dx < \infty .$
In this work, we use the Gaussian kernel $\frac{1}{\sqrt{2\pi \sigma^2}}\exp(-(x-m_x)^2/2\sigma^2)$ with $\sigma=.01$. Note that,
choosing other types of kernels does not change the result significantly. But the choice of the parameter value is
important \cite{BanerjeeJost2009,BanerjeeJost2007}. For small value of the parameter, the density plot  contains many random fluctuations and obscures the global features of the network structure. Where as for large value of the parameter, the details become very blurred. Thus, an optimum value $0.1$ of $\sigma$ is chosen from a range.
We compute the spectral density $f^*(x)=\frac{f(x)}{\int f(y)dy}.$

Now, we use the structural distance $D(G_1,G_2)$ between two different networks $G_1$ and $G_2$ as:
\bel{spec-dist}
D(G_1,G_2)=\sqrt{JS(f_1^*,f_2^*)},
\qe
where $JS(f_1^*,f_2^*)$ is the Jensen-Shannon  divergence measure between the spectral densities (of normalized graph Laplacian)
$f_1^*$ and $f_2^*$  of the networks $G_1$ and $G_2$, respectively \cite{banerjee12}.
Jensen-Shannon  divergence measure for two probability distributions $p_1$ and $p_2$ is defined as
 \bel{js}
 JS(p_1,p_2)=\frac{1}{2}KL(p_1,p)+\frac{1}{2}KL(p_2,p); \text{ where }p=\frac{1}{2}(p_1+p_2),
 \qe
where $KL(p_1,p_2)$ is the Kullback-Leibler divergence measure for two probability distributions $p_1$ and $p_2$  of a discrete 
random variable $X$ is defined as
\bel{kl}
KL(p_1,p_2)=\sum_{x\in X}p_1(x)\log\frac{p_1(x)}{p_2(x)}.
\qe
Though, theoretically, there exist isospectral graphs but they are very rare in real networks. Moreover, the  isospectral 
graphs are qualitatively quite similar in most respects. 

Now, we construct a distance table for all organisms studied here by measuring the distance $D(G_1, G_2)$ (defined in (\ref{spec-dist})) 
between  every pair of corresponding metabolic networks $(G_1, G_2)$. The table is used to 
produce a splits network \cite{Huson1998}, that not only shows different clusters among species, but also reflects the phylogenetic signal in the distance matrix.
We use the software package SplitsTree4 (version 4.13.1) \cite{Huson2006} for the  construction of splits (network) tree and neighbour-joining  tree.
To compare our results, another phylogenetic tree is also produced from SSU rRNA sequences of 79 species from NCBI: (\url{http://www.ncbi.nlm.nih.gov}) and using the software package FigTree v1.4.1 (\url{http://tree.bio.ed.ac.uk/software/figtree}).
\subsection{Data Acquisition and Processing}
We have downloaded the list of  enzymes of the 
79 completely sequenced organisms (7 eukaryotes, 13 archaea, 7 $\alpha$-subdivision of proteobacteria, 3 $\beta$-subdivision of
proteobacteria, 13 $\gamma$-subdivision of proteobacteria, 3 $\delta$-subdivision of proteobacteria, 17 low GC content gram positive bacteria,
3 high GC content gram positive bacteria, 1 fusobacteria, 5 chlamydia, 2 spirochete, 2 cyanobacteria, 1 radioresistant, 2 hyperthermophilic bacteria)
 from regularly updated {\it KEGG LIGAND} database (\url{http://www.kegg.jp/kegg/kegg2.html})\cite{Kanehisa06,Kanehisa14}.
The recent bioreaction database created by Michael et al. \cite{stelzer11}, which is an updated version of the same constructed
by Ma and Zeng \cite{ma03}, has been used to create the metabolic reaction database for each 79 species.
Michael et al. \cite{stelzer11} have considered several literatures  to decide about reversibility of each metabolic reactions  
in the dataset. They have  excluded currency metabolites to make the data more physiological meaningful.
\section{Results and Discussion}
\begin{figure}[!] 
\hspace{-3cm}
\includegraphics[width=1.6\textwidth]{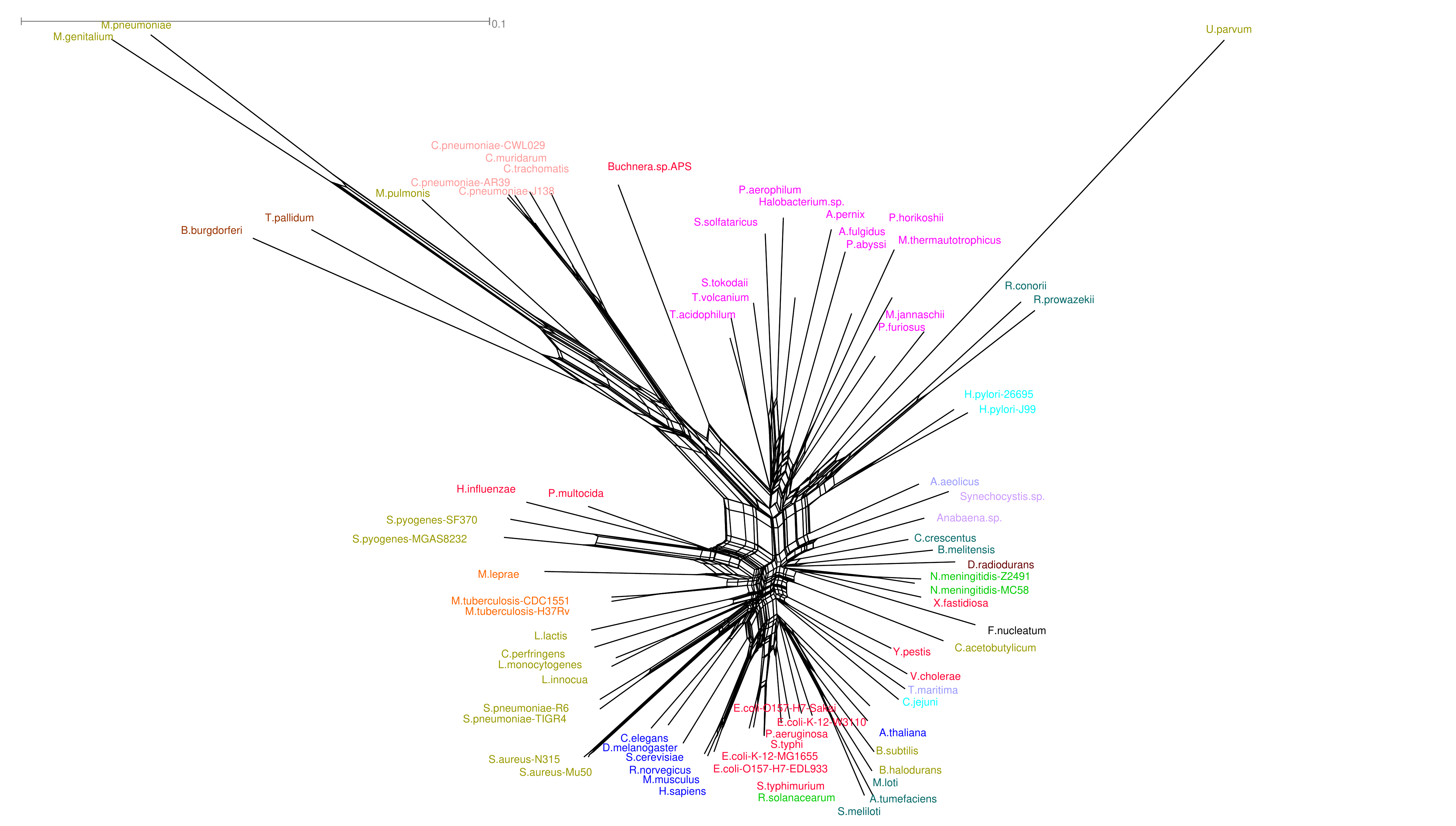}
\caption{The splits (network) tree constructed from the structural distances between metabolic-centric network of 79 species. {\color[rgb]{0,0,1}
$\blacksquare$} Eukaryotes; {\color[rgb]{1,0,1} $\blacksquare$} Archaea; and rest of are different subdivision of Bacteria;
{\color[rgb]{0,0.4,0.4} $\blacksquare$} $\alpha$-subdivision of proteobacteria, {\color[rgb]{0,0.8,0} $\blacksquare$} $\beta$-subdivision of proteobacteria,
{\color[rgb]{1,0,0.2} $\blacksquare$} $\gamma$-subdivision of proteobacteria, {\color[rgb]{0,1,1} $\blacksquare$} $\delta$-subdivision of proteobacteria,
{\color[rgb]{0.6,0.6,0} $\blacksquare$} Low GC content gram positive bacteria, {\color[rgb]{1,0.4,0} $\blacksquare$} High GC content gram positive bacteria,
{\color[rgb]{0,0,0} $\blacksquare$} Fusobacteria, {\color[rgb]{1,0.6,0.6} $\blacksquare$} Chlamydia, {\color[rgb]{0.6,0.2,0} $\blacksquare$} Spirochete, 
{\color[rgb]{0.8,0.6,1} $\blacksquare$} Cyanobacteria, {\color[rgb]{0.4,0,0} $\blacksquare$} Radioresistant bacteria,
{\color[rgb]{0.6,0.6,1} $\blacksquare$} Hyperthermophilic bacteria.} 
\label{metabolic-centric-network-pic}
\end{figure}

\begin{figure}[!] 
\hspace{-3cm}
\includegraphics[width=1.5\textwidth]{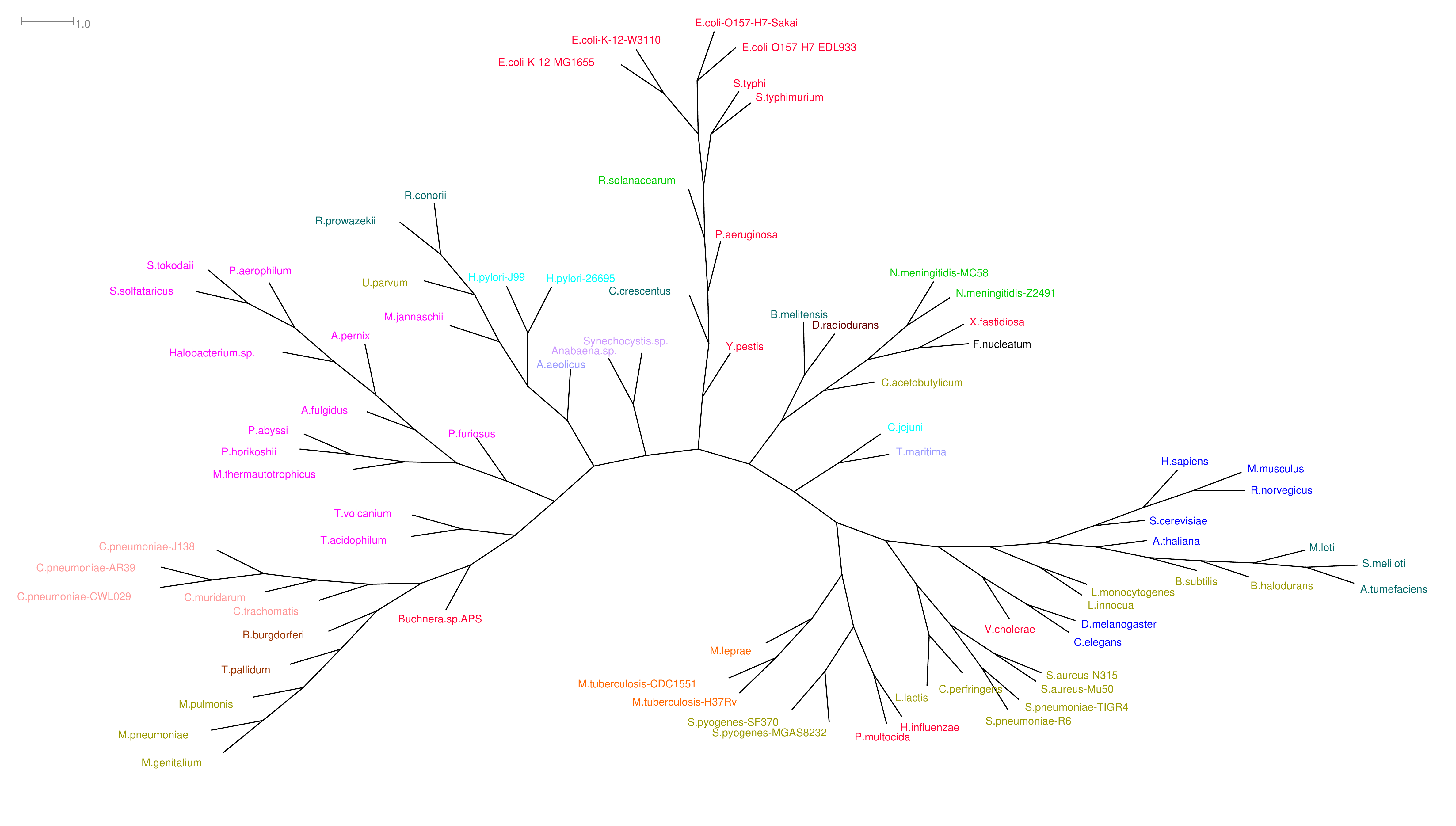}
\caption{The neighbour-joining tree constructed from the structural distances between metabolic-centric network of 79 species. {\color[rgb]{0,0,1}
$\blacksquare$} Eukaryotes; {\color[rgb]{1,0,1} $\blacksquare$} Archaea; and rest of are different subdivision of Bacteria;
{\color[rgb]{0,0.4,0.4} $\blacksquare$} $\alpha$-subdivision of proteobacteria, {\color[rgb]{0,0.8,0} $\blacksquare$} $\beta$-subdivision of proteobacteria,
{\color[rgb]{1,0,0.2} $\blacksquare$} $\gamma$-subdivision of proteobacteria, {\color[rgb]{0,1,1} $\blacksquare$} $\delta$-subdivision of proteobacteria,
{\color[rgb]{0.6,0.6,0} $\blacksquare$} Low GC content gram positive bacteria, {\color[rgb]{1,0.4,0} $\blacksquare$} High GC content gram positive bacteria,
{\color[rgb]{0,0,0} $\blacksquare$} Fusobacteria, {\color[rgb]{1,0.6,0.6} $\blacksquare$} Chlamydia, {\color[rgb]{0.6,0.2,0} $\blacksquare$} Spirochete, 
{\color[rgb]{0.8,0.6,1} $\blacksquare$} Cyanobacteria, {\color[rgb]{0.4,0,0} $\blacksquare$} Radioresistant bacteria,
{\color[rgb]{0.6,0.6,1} $\blacksquare$} Hyperthermophilic bacteria.} 
\label{metabolic-centric-network-nj-pic}
\end{figure}

\begin{figure}[!] 
\vspace{-1cm}
\hspace{-9cm}
\includegraphics[width=2.5\textwidth]{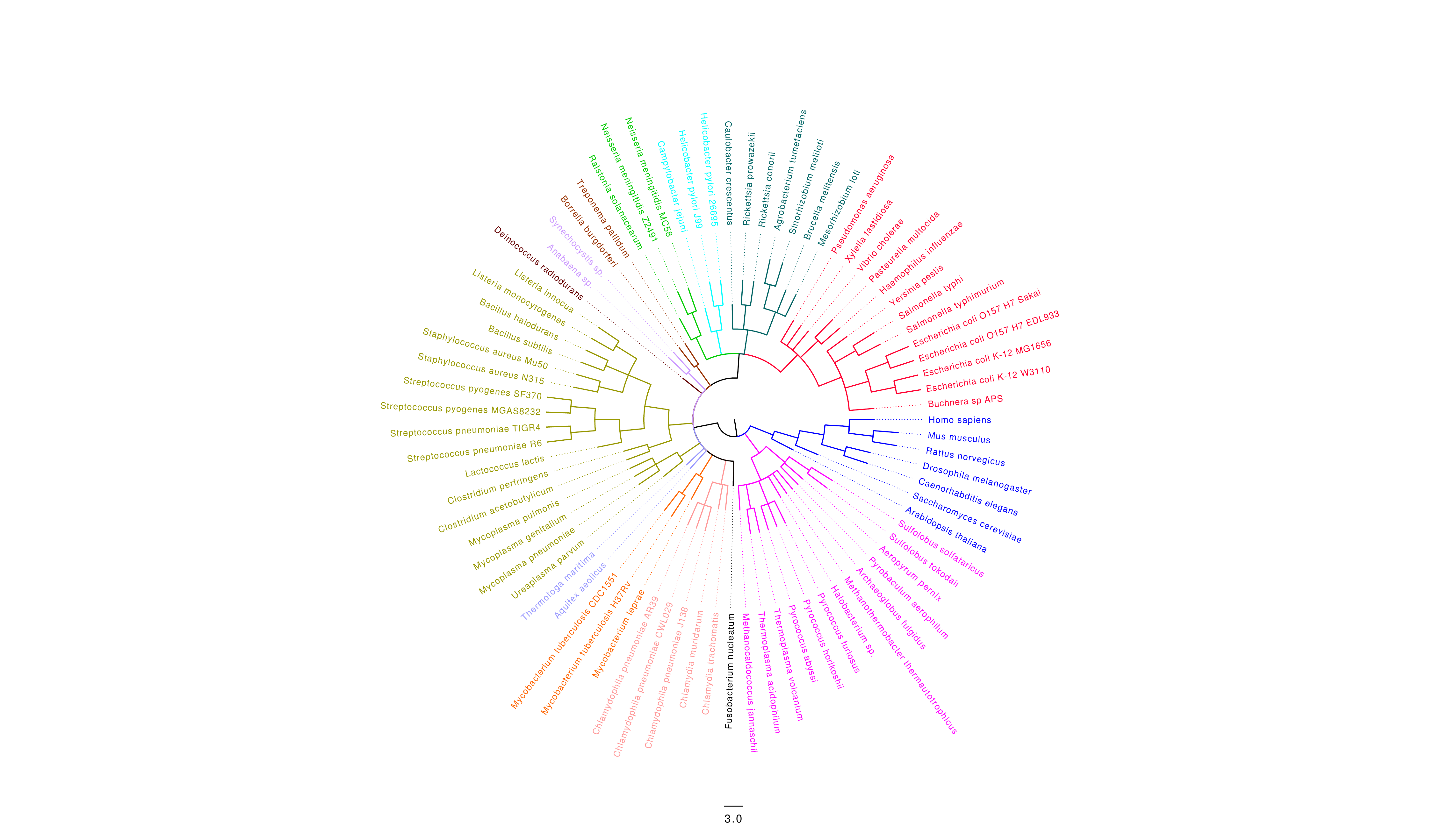}
\caption{The phylogenetic tree produced by SSU rRNA sequences of 79 species. {\color[rgb]{0,0,1}
$\blacksquare$} Eukaryotes; {\color[rgb]{1,0,1} $\blacksquare$} Archaea; and rest of are different subdivision of Bacteria;
{\color[rgb]{0,0.4,0.4} $\blacksquare$} $\alpha$-subdivision of proteobacteria, {\color[rgb]{0,0.8,0} $\blacksquare$} $\beta$-subdivision of proteobacteria,
{\color[rgb]{1,0,0.2} $\blacksquare$} $\gamma$-subdivision of proteobacteria, {\color[rgb]{0,1,1} $\blacksquare$} $\delta$-subdivision of proteobacteria,
{\color[rgb]{0.6,0.6,0} $\blacksquare$} Low GC content gram positive bacteria, {\color[rgb]{1,0.4,0} $\blacksquare$} High GC content gram positive bacteria,
{\color[rgb]{0,0,0} $\blacksquare$} Fusobacteria, {\color[rgb]{1,0.6,0.6} $\blacksquare$} Chlamydia, {\color[rgb]{0.6,0.2,0} $\blacksquare$} Spirochete, 
{\color[rgb]{0.8,0.6,1} $\blacksquare$} Cyanobacteria, {\color[rgb]{0.4,0,0} $\blacksquare$} Radioresistant bacteria,
{\color[rgb]{0.6,0.6,1} $\blacksquare$} Hyperthermophilic bacteria.}
 \label{16sRNA-pic}
\end{figure}

We construct metabolic networks separately for 79 species. Metabolites are  considered as nodes of the network and we connect two 
metabolites, a substrate and a product of a reaction, by an edge. 
All constructed networks are  not connected, rather they are composed with a giant  component and many isolated small components. In our work,
we only consider the giant component. We believe, this part of the network is composed with the most studied metabolic pathways, thus it may
be extensively revised and contains more errorless information. Here, we also consider the underlying undirected graph which itself carries
adequate structural information  for our work. This method can be easily extended to study a directed network.
Now, we use our method to construct a distance table for metabolic networks of all the organisms studied here. This table is used  to produce a splits network which can extract phylogenetic signals that are missed in other tree-representations. The splits network shows that the data contained in that matrix has a substantial amount of phylogenetic signal and some parts of the data are tree-like (see \cite{Huson1998} for details). We also construct a tree by using neighbour-joining method from our distance matrix to visually capture and compare different clusters easily. 
It is our goal is not to reproduce the phylogenetic tree, but to cluster the species by quantifying the similarity in their metabolic network structures. 
Now, we analyze the different clusters in the splits network (figure(\ref{metabolic-centric-network-pic})) and  neighbour-joining tree (figure(\ref{metabolic-centric-network-nj-pic})) produced from the distances
between the spectral densities (of normalized graph Laplacian) of metabolic networks of 79 organisms studied here\footnote{Both of the trees (figure(\ref{metabolic-centric-network-pic}) and figure(\ref{metabolic-centric-network-nj-pic})) do not demonstrate any significant difference in clustering.}. To compare the clustering from molecular phylogeny, another tree (figure(\ref{16sRNA-pic})) is produced from SSU rRNA sequences, which are highly conserved, of those 79 species. 
On one hand, splits network reflects the evolutionary signature in the data, but on the other hand, some of the clusters (in splits-network and the neighbour-joining tree) are different from the same in the tree constructed by SSU rRNA sequences (Robinson-Foulds distance between these two trees is 75). This suggests that two species which cluster together in spite of belonging to different classes, that is, in different clusters in the phylogenetic tree (figure(\ref{16sRNA-pic})) produced by SSU rRNA sequences of 79 species, may have  evolved in similar environmental condition or with similar evolutionary histories that make their metabolic networks topologically more similar than others. Now, we explore the commonalities between these species.
 
In our analysis, we observe that {\it Mesorhizobium loti} and {\it Sinorhizobium meliloti}, which are two symbiotes, cluster with a
 free-living (non-symbiote) species {\it Agrobacterium tumefaciens} of the same group, $\alpha$-subdivision of proteobacteria. This may be
 explained by the genome similarity between  {\it S. meliloti} and {\it A. tumefaciens} which suggests convergent evolution (see \cite{wood01}
 for more details). Moreover, all these organisms are associated with the roots of  crop plants \cite{rudra08}.
 
 All the chlamydia species strongly cluster together which is also observed in \cite{ma04}.
 One $\gamma$-proteobacteria {\it Buchnera sp.~APS} appears near the branch of chlamydia.
{\it Buchnera sp.~APS} is endocellular symbiote of aphids. Like parasites, {\it Buchnera sp.~APS} has reduced its genome size and it depends upon host for that of the essential amino acids and vitamins \cite{shigenobu00}. It reflects that its lifestyle is similar to parasite. 

Most of the parasites, from different groups, cluster together. Where-as the remaining three parasites, {\it Rickettsia prowazekii}, {\it Rickettsia conoii}
 from $\alpha$-subdivision of proteobacteria and {\it Ureaplasma parvum} which is a low GC containing gram  positive bacterium, 
 form a separate branch in our tree. In between these two parasitic branches all archaea cluster together. In the previous 
 studies, \cite{podani01} and \cite{ma04} which have a  slightly different observation, all the parasites cluster together. 

 Most of the species from $\gamma$-subdivision of proteobacteria including 
 {\it Salmonella typhi, Salmonella typhimurium,
 Pseudomonas aeruginosa} and 
  four very related strains of {\it Escherichia coli}
  form strongly supported cluster with a lethal plant
 pathogen from $\beta$-subdivision of proteobacteria, {\it Ralstonia solanacearum}. 
 In whole-genome alignment,  {\it P. aeruginosa} is congruous with {\it R. solanacearum} \cite{arodz08}.  

 Moreover, like in other methods, all the three high GC content gram positive bacteria cluster together in our tree.
 
\begin{figure}[!] 
\hspace{-4cm}
\includegraphics[width=1.6\textwidth]{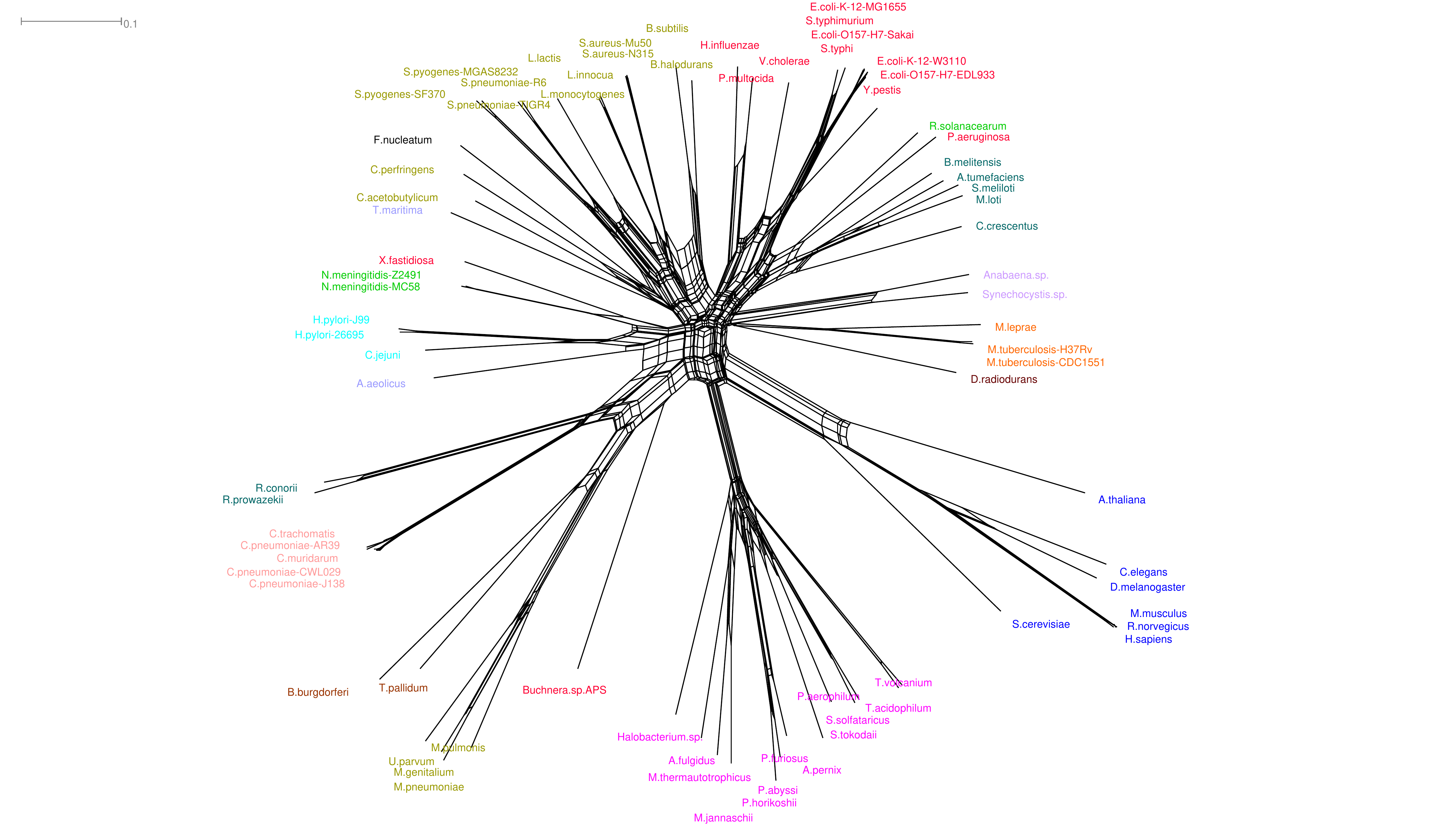}
\caption{The splits (network) tree constructed from enzymes content of genome-based metabolic networks of 79 organisms with evolution distance
based on Jaccard index. {\color[rgb]{0,0,1}
$\blacksquare$} Eukaryotes; {\color[rgb]{1,0,1} $\blacksquare$} Archaea; and rest of are different subdivision of Bacteria;
{\color[rgb]{0,0.4,0.4} $\blacksquare$} $\alpha$-subdivision of proteobacteria, {\color[rgb]{0,0.8,0} $\blacksquare$} $\beta$-subdivision of proteobacteria,
{\color[rgb]{1,0,0.2} $\blacksquare$} $\gamma$-subdivision of proteobacteria, {\color[rgb]{0,1,1} $\blacksquare$} $\delta$-subdivision of proteobacteria,
{\color[rgb]{0.6,0.6,0} $\blacksquare$} Low GC content gram positive bacteria, {\color[rgb]{1,0.4,0} $\blacksquare$} High GC content gram positive bacteria,
{\color[rgb]{0,0,0} $\blacksquare$} Fusobacteria, {\color[rgb]{1,0.6,0.6} $\blacksquare$} Chlamydia, {\color[rgb]{0.6,0.2,0} $\blacksquare$} Spirochete, 
{\color[rgb]{0.8,0.6,1} $\blacksquare$} Cyanobacteria, {\color[rgb]{0.4,0,0} $\blacksquare$} Radioresistant bacteria,
{\color[rgb]{0.6,0.6,1} $\blacksquare$} Hyperthermophilic bacteria.}
 \label{enzymes-content-metabolic-network-pic}
\end{figure}

\begin{figure}[!] 
\hspace{-3cm}
\includegraphics[width=1.5\textwidth]{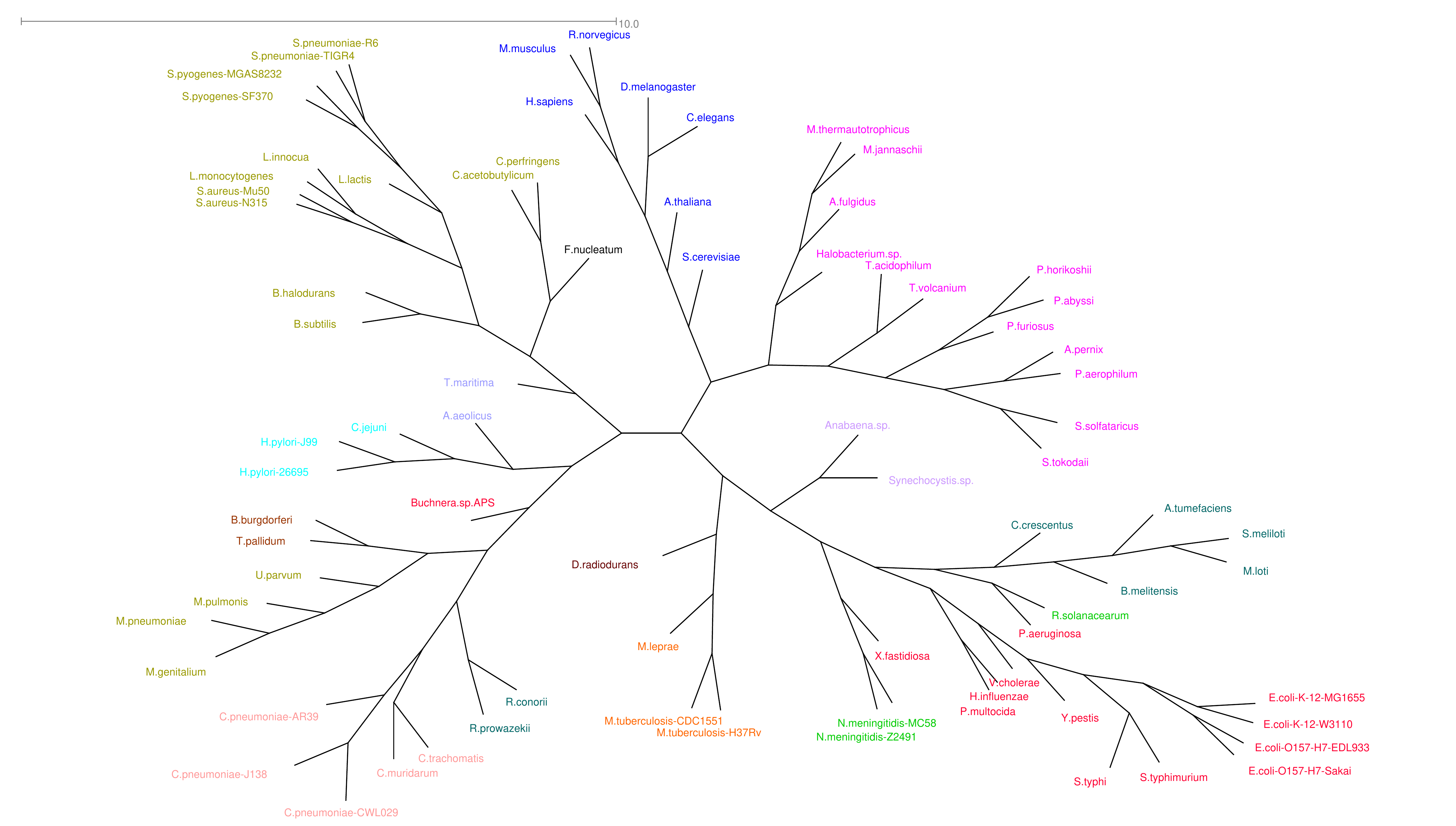}
\caption{The neighbour-joining tree constructed from enzymes content of genome-based metabolic networks of 79 organisms with evolution distance
based on Jaccard index. {\color[rgb]{0,0,1}
$\blacksquare$} Eukaryotes; {\color[rgb]{1,0,1} $\blacksquare$} Archaea; and rest of are different subdivision of Bacteria;
{\color[rgb]{0,0.4,0.4} $\blacksquare$} $\alpha$-subdivision of proteobacteria, {\color[rgb]{0,0.8,0} $\blacksquare$} $\beta$-subdivision of proteobacteria,
{\color[rgb]{1,0,0.2} $\blacksquare$} $\gamma$-subdivision of proteobacteria, {\color[rgb]{0,1,1} $\blacksquare$} $\delta$-subdivision of proteobacteria,
{\color[rgb]{0.6,0.6,0} $\blacksquare$} Low GC content gram positive bacteria, {\color[rgb]{1,0.4,0} $\blacksquare$} High GC content gram positive bacteria,
{\color[rgb]{0,0,0} $\blacksquare$} Fusobacteria, {\color[rgb]{1,0.6,0.6} $\blacksquare$} Chlamydia, {\color[rgb]{0.6,0.2,0} $\blacksquare$} Spirochete, 
{\color[rgb]{0.8,0.6,1} $\blacksquare$} Cyanobacteria, {\color[rgb]{0.4,0,0} $\blacksquare$} Radioresistant bacteria,
{\color[rgb]{0.6,0.6,1} $\blacksquare$} Hyperthermophilic bacteria.}
 \label{enzymes-content-metabolic-network-nj-pic}
\end{figure}

  All the above results can also be seen in the tree (splits network and neighbour-joining tree) generated by another method  based on a dissimilarity measure  using Jaccard-index \cite{ma04} on the presence of enzymes in 79 species (see figures (\ref{enzymes-content-metabolic-network-pic}) and (\ref{enzymes-content-metabolic-network-nj-pic})).
 Now, we discuss our findings which were not captured by most of the other methods.
 
 Interestingly, our result shows the close relationship between bacteria and eukaryotes that could not captured by the methods used 
in  \cite{podani01} and \cite{ma04}. This may happen as most of the metabolic enzymes of eukaryotes are of bacterial origin \cite{rivera98}.   
 
 Six species, from different phylogenetic groups, with the similar life histories form a  group consisting of a plant in eukaryotes {\it Arabidopsis thaliana}, 
 two low GC content gram positive bacteria {\it Bacillus subtilis} and
  {\it Bacillus halodurans}, three  $\alpha$-subdivision of proteobacteria {\it Mesorhizobium loti},
  {\it Sinorhizobium meliloti} and {\it Agrobacterium tumefaciens}. 
   {\it B. subtilis}, which is taxonomically related
  to another alkaliphilic bacterium {\it B. halodurans}, is a free-living low GC content gram positive bacterium and associated with the 
  root of the plant {\it A. thaliana}  \cite{hideto00}. As we know that {\it M. loti} and {\it S. meliloti} are nitrogen
  fixation  symbiotic bacteria living at the root surface of leguminous plant. 
 But {\it A. tumefaciens}, which  is a free-living pathogenic bacterium, is found in soil and forms biofilm 
 after attaching to the host root \cite{danh07, rudra08}. 
 
 A hyperthermophilic marine bacterium {\it Aquifex aeolicus} clusters with two cyanobacteria {\it Synechocystis} and {\it Anabaena}. All of them participate in nitrogen-fixing \cite{Makarova,Studholme,Tamagnini}.
 
 Moreover, our analysis shows that  the four species, which are free-living human pathogens that cause respiratory diseases and are 
from three different groups, appear close to each other in our splits (network) tree. These species are: two gamma proteobacteria 
{\it Haemophilus influenzae} and {\it Pasteurella multocida}, one 
 actinobacterium {\it Mycobacterium tuberculosis}, and a low GC content gram positive bacterium {\it Streptococcus pneumoniae}. 
 They infect the upper respiratory tract of human. But, another low GC content gram positive bacterium  {\it Mycoplasma pneumoniae}
 which also infects the upper respiratory tract as well as the lower respiratory  tract  clusters with the parasitic branch.      
 
 Evidence shows that the horizontal gene transfer took place between the radioresistant  {\it Deinococcus radiodurans} and gamma 
 proteobacteria {\it Vibrio cholerae} long ago \cite{eisen00}. They  are not very far from each other in our split tree. 

 Not all the clusters traced by our method are meaningful, for example,  one plant pathogen {\it Xylella fastidiosa} from $\gamma$-subdivision of proteobacteria clusters with {\it Neisseria meningitidis MC58, Neisseria meningitidis Z2491} which are  $\beta$-subdivision of proteobacteria. The locations of all the clusters of low GC content gram positive bacteria  and $\gamma$-subdivision of proteobacteria could not be justified.
However, many clusters, between various species, found by our study show interesting similarities in some aspects of their lifestyle.
\section{Conclusion}
Here we have used the spectrum of normalized graph Laplacian which reveals global as well as  local architectures, which have emerged from 
the evolutionary process like motif duplication or joining, random rewiring, random edge deletion, etc., of a network. A structural 
difference measure, based on the divergence between  two spectral densities, has been applied to find the topological distances between
the metabolic networks of 79 species. A splits (network) tree and a neighbour-joining tree have been used to explore the evolutionary relation between these networks. Our study shows some 
new interesting insights into the phylogeny of different species, constructed on the basis of their metabolic networks.

The spectrum of non-normalized Laplacian matrix or adjacency matrix can also be used for the same study, but a correlation has been observed 
between the degree distribution of a network and the spectral density of these matrices  (see \cite{DorogovtsevEtAl2004,ZhanEtAl2010}).
Due to irregular structure and different sizes, it is difficult to compare different real networks. To investigate the structural similarities,
graphs of similar sizes can be aligned to each other. Since for all the networks, the spectrum of the normalized graph Laplacian are bound 
within a specific range ($0$ to $2$), we have an added advantage for comparing spectral plots of  networks with different sizes.
\section{Acknowledgments}
 KD gratefully acknowledge the financial support from {\it CSIR} (file number 09/921 (0070)/2012-EMR-I), Government of India.  The part of the  work done by BD is financially supported by the project: National Network for Mathematical and Computational Biology (SERB/F/4931/2013-14)  funded by {\it SERB}, Government of India.
The authors are thankful to Ravi Kumar Singh for tree construction. Authors are thankful to Partho Sarothi Ray for helping to prepare the manuscript.


\pagebreak

\end{document}